\documentclass{aastex631}

\usepackage{natbib}
\usepackage{orcidlink}
\usepackage{amsmath}

\shorttitle{Observed Rate Variation in Superflaring G-type Stars}
\shortauthors{Crowley}

\graphicspath{{./}{figures/}}

\begin{document}

\title{Observed Rate Variations in Superflaring G-type Stars}

\author{James Crowley}
\affiliation{Sydney Institute for Astronomy, University of Sydney, NSW 2006, Australia}

\author{Michael S. Wheatland \orcidlink{0000-0001-5100-2354}}
\affiliation{Sydney Institute for Astronomy, University of Sydney, NSW 2006, Australia}

\author{Kai Yang \orcidlink{0000-0002-7663-7652}}
\affiliation{Sydney Institute for Astronomy, University of Sydney, NSW 2006, Australia}

\begin{abstract}
Flare occurrence on the Sun is highly variable, exhibiting both short term variation due to the emergence and evolution of active regions, and long-term variation from the solar cycle. 
On solar-like stars, much larger stellar flares (superflares) have been observed, and it is of interest to determine whether observed rates of superflare occurrence exhibit similar variability to solar flares.
We analyse 274 G-type stars using data from the Transiting Exoplanet Survey Satellite (TESS) and identify seven stars which exhibit statistically significant changes in the rate of superflare occurrence by fitting a piecewise constant-rate model with the Bayesian Blocks algorithm \citep{10.1088/0004-637X/764/2/167}. We investigate the properties of these stars and their flaring rates, and discuss the possible reasons for the low number of stars with detectable rate variation.
\end{abstract}


\section{Introduction} \label{sec:intro}

Flares are rapid, explosive releases of magnetic energy on stars resulting from reconnection events in a star's magnetic field \citep{pettersen_1989}. The reconnection results in a change in the field line connectivity and the release of magnetic energy in a variety of forms, ranging from transient emissions across the electromagnetic spectrum and particle acceleration to plasma heating and bulk movement \citep{10.1146/annurev-astro-082708-101757}. On the Sun, solar flares have been observed to release $\lesssim 10^{32}$ erg over time scales of minutes to hours. Observations of other stars, through surveys from the \textit{Kepler} \citep{Kepler_info} and \textit{TESS} \citep{TESS_info} satellites have revealed the existence of much larger flaring events dubbed `superflares' \citep{10.1086/308325} with bolometric energies from $10^{33}$ erg to as high as $10^{37}$ erg \citep{10.3847/0004-637X/829/1/23}. Superflares have been observed to occur not only on fast-rotating solar-type stars \citep{10.1088/0067-0049/209/1/5, 10.1038/nature11063}, but also on slowly-rotating solar-type stars with orbital periods similar to the Sun \citep{10.3847/1538-4357/ab14e6}. 

Solar flares act as drivers of near-Earth space weather through the release of highly energetic particles and radiation, and through the coronal mass ejections (CMEs) that often accompany larger flares \citep{10.1016/j.asr.2008.11.004}. 
The radiation and energetic particles from both flares and CMEs can lead to damage of satellite electronics, disruption of high-frequency radio transmissions, and failure of power grids due to geomagnetically-induced currents \citep{https://doi.org/10.1029/2019SW002278}. 
The first observation of a solar flare \citep{10.1093/mnras/20.1.13} resulted in a geomagnetic storm on Earth which led to widespread damage to the telegraph system of North America and Europe \citep{10.1016/j.asr.2005.12.021, 10.1016/j.asr.2006.01.013}. This event is now acknowledged to be one of the largest solar flares ever observed, with a bolometric energy  $\approx 5\times 10^{32}$ erg \citep{10.1051/swsc/2013053}. 
In recent years, large solar flares have disrupted communications and power systems at Earth; most notably the March 1989 and Halloween 2003 flares. Geomagnetic storms from these flares/CMEs saw communication blackouts as satellites were damaged or disrupted, as well as outages due to damage to transformers in power systems \citep{10.1016/S1364-6826(02)00128-1, 10.1016/S1364-6826(02)00036-6,balch2004intense, 10.1051/swsc/2012004}. Previous solar storms have resulted in health complications in astronauts \citep{balch2004intense}, and estimates of the damage resulting from a Carrington-level event are in the range of trillions of dollars (see e.g. \citealt{10.1111/risa.12765, 10.1007/s11214-017-0456-3} and references therein).

Stellar superflares, which may be $10^4$ times as energetic as the largest solar flares, can adversely affect the habitability of nearby planets. Studies into the effects of single \citep{10.1089/ast.2009.0376} and multiple \citep{10.1089/ast.2017.1794} superflares on an Earth-like planet orbiting an active M dwarf demonstrate rapid depletion in atmospheric ozone concentration, resulting in high surface UV flux. Studies also suggest that superflares may play a role in the development of life through the enhancement of prebiotic chemistry \citep{10.1017/S174392131600226X}. 

Various studies have focused on stars similar to the Sun to estimate the occurrence frequency of superflares on solar analogues (e.g. \citet{10.1088/0067-0049/209/1/5, 10.3847/1538-4357/ab6606}). \citet{10.1186/s40623-015-0217-z} found that $10^{33}$ erg superflares occur on solar-type stars about once every 500 years. \citet{10.3847/1538-4357/ab14e6} and later \citet{10.3847/1538-4357/abc8f5} analysed slower rotating solar-type stars similar to the Sun, with \citet{10.3847/1538-4357/abc8f5} estimating that a superflare releasing $10^{34}$ erg could occur once every $\approx 6000$ years. However, these estimates are based just on the number of observed flares, the number of observed stars, and the observation time. It is unclear whether the Sun is capable of producing flares this large~\citep{10.1093/pasj/65.3.49}.

The statistics of solar flare occurrence are well established, and are generally characterised using a flare frequency-energy distribution (FFD) which describes the frequency of flares per unit energy. Typically, this distribution is fit over a range of energies by a power-law distribution
\begin{equation}
    N(E) = AE^{-\gamma},
\end{equation}
or equivalently,
\begin{equation}\label{eq:N(E)}
    N(E) = \lambda_0 (\gamma -1)E_{0}^{\gamma -1} E^{-\gamma}
\end{equation}
where $\gamma$ is the power-law index and $\lambda_0$ describes the rate of flaring above the energy threshold $E_0$ \citep{10.1023/A:1012749706764}. Similar power-law distributions are observed for related quantities such as the peak flux of flares in different wavelengths.
The power-law index varies depending on the specific quantity. 
Hard X-ray measurements of the peak flux of solar flares give a power-law index of $\gamma = 1.73 \pm 0.07$ \citep{10.1007/s11214-014-0054-6} which notably holds over several orders of magnitude, from nanoflares (energies in the range of $\approx 10^{24}$ erg, \citealp{10.1086/308867, 2022arXiv220312484A}) up to X-class flares ($\approx 10^{31}$ erg, \citealp{ 10.1007/978-3-642-15001-2, 2022arXiv220312484A}). For the bolometric energy of solar flares, the power-law index is around $1.62\pm 0.12$ \citep{10.1007/s11214-014-0054-6}. 
In stellar superflares the FFD is measured in terms of the bolometric energy of flares and is also typically fitted with a power-law distribution. The power-law fits for stellar superflares have been observed to have similar indices to the power-law fits for solar flares \citep{10.1088/0067-0049/209/1/5, 10.1186/s40623-015-0217-z, 10.3847/0004-637X/829/1/23, 10.3847/1538-4365/abda3c}. 

A second distribution used to characterise solar flare occurrence is the waiting time distribution (WTD), i.e.\ the distribution of times between flares.
If flares occur randomly in time with a constant rate $\lambda_0$, then the WTD will have the Poisson form
\begin{equation}
    P(\Delta t) = \lambda_0 {\rm e}^{-\lambda_0 \Delta t}.
\end{equation}    

Studies of flaring in individual active regions on the Sun show that flare occurrence can often be described as a constant-rate Poisson process, at least for short periods of observation \citep{10.1023/A:1012749706764}, and in that case the WTD is approximately exponential. For all active regions on the Sun, the overall WTD corresponds to the superposition of events from different active regions, and is also approximately exponential over short timescales (i.e.\ the rate of flaring is approximately constant, and corresponds to the total rate of events in different regions). Rate variation is commonly observed over longer timescales. In this case the WTD can be approximated with a piecewise-constant rate model \citep{2000ApJ...536L.109W,10.1023/A:1022430308641}, and has the form:

\begin{equation}\label{eq:piecewise-poisson}
P(\Delta t) = \sum_i \frac{n_i}{N}\lambda_i {\rm e}^{-\lambda_i\Delta t},
\end{equation}
where $n_i=\lambda_i t_i$ is the number of events in the interval $t_i$, when the rate is $\lambda_i$, and $N=\sum_i n_i$ is the total number of events. 
For a range of distributions of $\lambda_i$, the resulting WTD can exhibit a power-law tail $P(\Delta t) \propto (\Delta t)^{-p}$ \citep{2000ApJ...536L.109W,10.1007/978-3-642-15001-2}. 
Observations of soft X-ray flare events from the whole Sun over decades show a significant change in the power-law index in the tail of the WTD, varying from $p=2.95\pm 0.5$ for years around solar maximum to $p=2.2\pm0.1$ for years around solar minimum \citep{10.1023/A:1022430308641}. These results illustrate the utility of the WTD for investigating variation in flaring rate.

Magnetic activity on the Sun -- magnetic phenomena such as solar flares and sunspots -- exhibits variability on both short and long time scales. 
Over short time scales, the rate of solar flare occurrence is primarily affected by the presence of sunspots, with almost all large flares being associated with large, magnetically complex $\delta$-spot regions \citep{10.1086/309303}. These sunspots also rotate onto and off the solar disk over several weeks due to the rotation of the Sun. Over longer time scales, magnetic activity on the Sun periodically fluctuates over 11 years with the solar cycle \citep{10.1146/annurev-astro-081913-040012}. Observations have shown that this occurs consistently between various forms of magnetic activity such as sunspot counts \citep{10.1023/A:1005007527816}, the occurrence rate of solar flares and the average energy of flares \citep{10.1023/A:1022430308641}, as well as through spectroscopic measurements of the emission lines in the chromosphere \citep{10.1086/151310}. 

On other stars, starspots have been inferred primarily through quasi-periodic brightness oscillations in the integrated light curves of the stars, which are interpreted as rotational modulation of the total luminosity due to spots~\citep{10.1088/0004-637X/771/2/127}. If starspots are required for stellar flares, then we might expect to see a dependence of stellar flare occurrence on rotational phase as a large starspot rotates around the surface of the star. However, measurements of the rotational phase of stellar flares have shown a uniform distribution with phase. This may indicate a lack of dependence of flaring on starspots, or flaring in multiple starspots across the whole stellar surface, or perhaps the presence of a polar starspot which would not result in rotational modulation due to always being in view \citep{10.1088/0004-637x/797/2/121, 10.1093/mnras/sty1963, 10.1093/mnras/stz2205, 10.1093/mnras/staa923}. To date there have not been any reports of short-term variability in the rate of occurrence of stellar superflares, and there is only weak evidence for long time-scale rate variability from activity cycles through measurement of the fractional luminosity emitted in flares \citep{10.3847/2515-5172/ab45a0}. Long time-scale variability is primarily observed in other tracers of magnetic activity, e.g.\ spectroscopic measurements of the chromosphere \citep{10.1051/0004-6361/201629518}. 
Spectroscopic surveys such as the Mount Wilson survey have identified activity cycles for a small, select sample of stars \citep{10.1086/307794}. Data from photometric surveys such as \textit{Kepler} have also been used to identify activity cycles: \citet{10.1051/0004-6361/201730599} measured short activity cycles up to 6 years in length through oscillations in the luminosity of the star, and \citet{10.1093/mnras/stz782} and \citet{10.3847/1538-4357/aaa026} combined data from \textit{Kepler} with spectroscopic and astroseismological measurements to measure an activity cycle of $7.41\pm 1.16$ years for the star KIC 8006161.

This project aims to identify whether there is short-term time variability in the rate of superflare occurrence, akin to the variability observed with solar flares. We focus on G-type stars which are the most similar to the Sun. To test for variations in the rate of superflare occurrence, we apply techniques previously used for solar flares \citep{10.1023/A:1022430308641}, namely the Bayesian Blocks algorithm \citep{10.1088/0004-637X/764/2/167} which optimally decomposes events in time into a piece-wise constant-rate model, and the waiting-time distribution, which, as shown by equation~(\ref{eq:piecewise-poisson}), can provide information on the rate distribution in time \citep{2000ApJ...536L.109W,10.1023/A:1022430308641} 
We also analyse the FFD for superflares, and compare the results with solar observations.

We have analysed 274 G-type stars using data from \textit{TESS} up to Sector 53 (concluding 2022-07-09). We describe the details of the data processing pipeline in Section \ref{sec:methods}, including the process of removing stellar variability, flare identification and then the statistical analysis. In Section \ref{sec:results}, we characterise the observed rate variation, and then we discuss the results, including the limitations of our analysis, in Section \ref{sec:discussion}.

\section{Methods} \label{sec:methods}

\subsection{Sample Selection}
Flare catalogues for highly flaring stars in the \textit{TESS} Input Catalog (TIC) are available, e.g.\ \citet{10.3847/1538-3881/ab5d3a, 10.3847/1538-4365/abda3c}. As we are attempting to identify G-type stars which exhibit flare rate variation, we select the most flare-producing stars from these previous studies. The effective temperature $T_{\text{eff}}$ and surface gravity $g$ are obtained from the TIC if available. If a star does not have parameters listed in the TIC, we fall back to using data from GAIA Data Release 2 and note these stars for inspection. We ensure each chosen star is a G-type star by requiring $5100$~K $\leq T_{\text{eff}} \leq 6000$~K and $\log g > 4.0$, consistent with the criteria used by \citet{10.3847/1538-4365/abda3c}.

\subsection{Processing and Quiescent Luminosity Modelling}\label{subsec:processing}
\textit{TESS} provides data in the form of a Target Pixel File (TPF), which encodes the flux of each pixel over time, or as Simple Aperture Photometry (SAP) data, an integrated light curve containing the flux from the pixels within a defined aperture mask. \textit{TESS} provides observations for around 20,000 stars per sector in 2-minute cadence, and following the start of the extended mission with Sector 27, also provides 20-second cadence for up to 1000 stars per sector \citep{TESS_extended_mission}. Each observation sector is viewed for approximately 27.4 days, divided into two segments corresponding to the ecliptic orbit of the satellite. 

For each target star, we download the TPFs in 2-minute cadence and perform a visual inspection to confirm the star is not a binary star. Then, data points which have quality flags \texttt{1, 2, 4, 8, 16, 32, 64, 128, 512, 4096, 16384} are removed from the data set. These flags correspond to systematic issues such as altitude tweaks and scattered light incidents, unexpected brightening events (e.g. argabrightening, cosmic rays), as well as impulsive outliers and anomalous data points \citep{TESS_data_products, TESS_data_products_sec20}. These events are undesirable and can affect the detection of superflares by introducing similar large, impulsive structures into the lightcurve, and hence they are removed. Finally, we create the integrated light curve of the star by integrating over each pixel specified in the provided aperture mask.

In order to identify flares, we first need to model the variability present in the light curve, which includes both systematic and astrophysical components. By doing this, we can construct a model of the quiescent luminosity of the star and separate the remaining outliers from this model such as flares. In \textit{TESS}, systematic non-physical variations are due to instrumental effects, including the differing efficiency of each camera. Astrophysical variability present in the light curve most commonly consists of rotational modulation -- quasi-periodic variation due to the presence of starspots \citep{10.3847/1538-4357/ab14e6}.

We begin by identifying the gaps in the data, which separate periods of continuous observation consisting of at least 720 consecutive data points with no gaps larger than an hour between any data points. The data in each of the continuous observation periods is normalised by dividing by the median value of the flux in that period. Large excursions from the mean are masked out before calculating the median via an iterative `sigma-clipping' process: values above 2.5 standard deviations away from the mean are masked out of the data set, over 10 repeated iterations. 
Long term systematic variation is then modelled by fitting a third-order polynomial to each continuous observation period. This polynomial fit is then subtracted from the light curve.
Using this systematics-corrected light curve, we identify the presence of periodicity (e.g.\ associated with rotational modulation) using a Lomb-Scargle periodogram. A star is considered to have periodicity if a significant signal is detected (Lomb-Scargle power $>$ 0.1).

\begin{figure}
\plotone{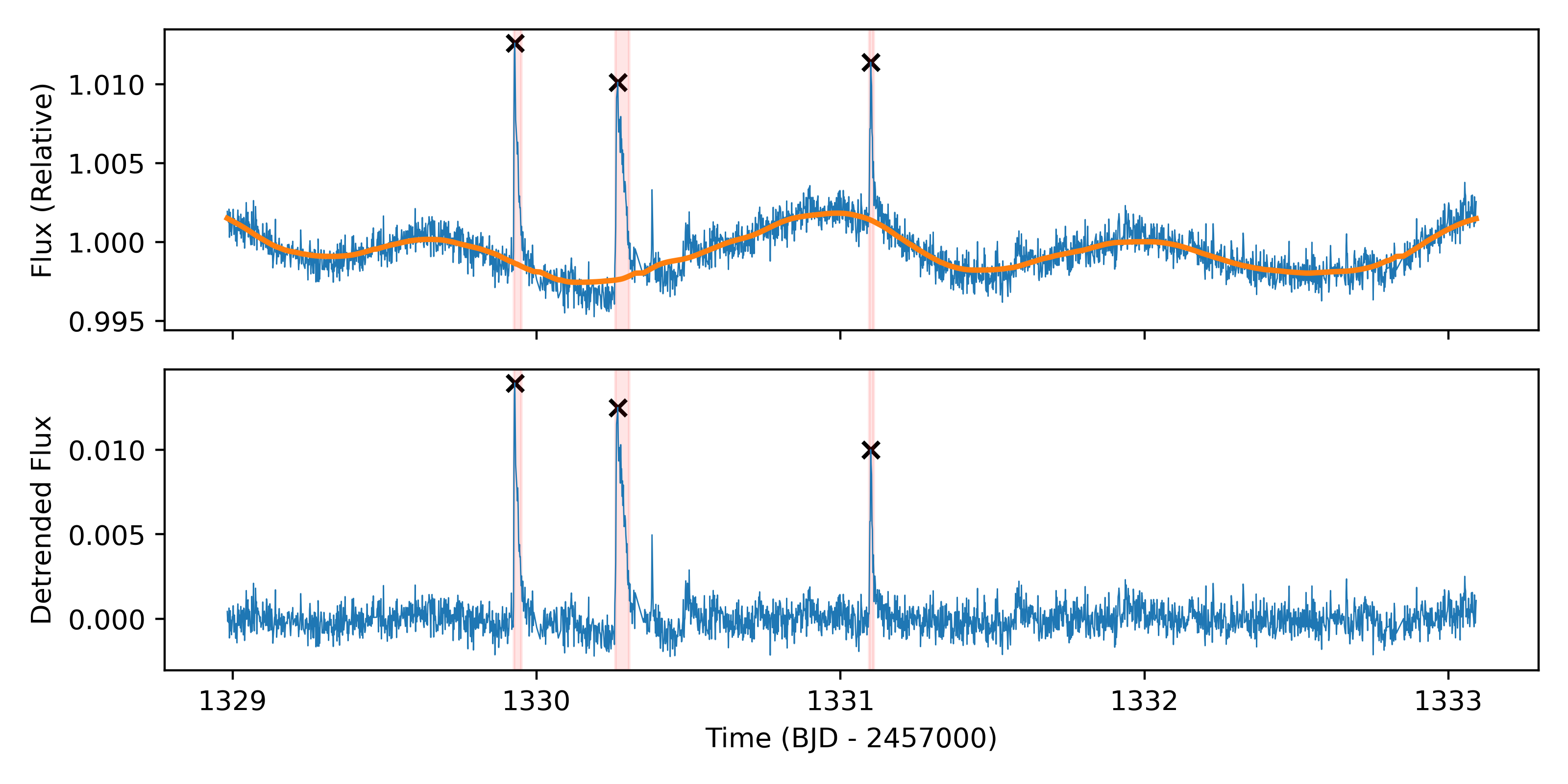}
\caption{
    \textbf{Top:} An excerpt of the light curve of TIC 364588501, showing the raw data (blue) and the background model calculated using the nonlinear iterative filtering technique (orange line). Several flare candidates are highlighted in red, with their peaks marked with crosses. 
    \textbf{Bottom:} The same light curve after the background model is subtracted.}
    \label{fig:detrending}
\end{figure}

To model the remaining variability, we use the nonlinear iterative filter previously used by \citet{10.1111/j.1365-2966.2004.07657.x} for processing light curves to identify exoplanet transits.  An example of the application of this filter is shown in Figure \ref{fig:detrending}. 
This filter has two parameters, \texttt{nmed} and \texttt{nlin}, which control the window sizes of the median filter and box-car filter used in the algorithm. From manual experimentation, we find that due to the variable length of flaring events (as opposed to the fixed time scales of exoplanet transit events), scaling the length of the median filter window by the detected period of the star generally results in an accurate model of the quiescent background, and avoids including parts of the flare in the background model. 
Specifically, we choose $\mathtt{nmed} = 100 \times T$ and $\mathtt{nlin} = 30$ for stars with period $T$, and otherwise use $\mathtt{nmed}=200$ for stars with no detected period. 

\subsection{Flare Finding}\label{subsec:flare-finding}

\begin{figure}[h]
\plotone{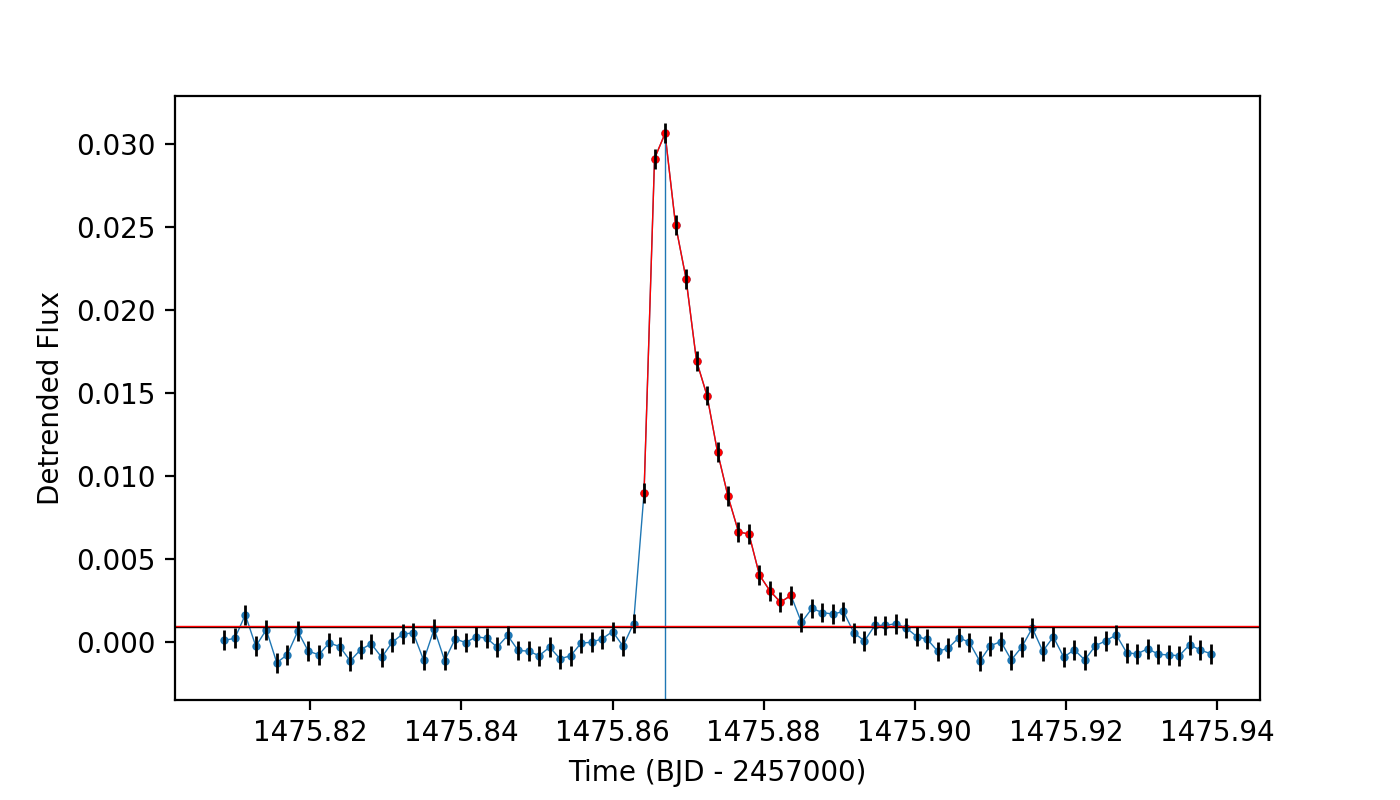}
\caption{
    A flare on TIC 364588501, depicting the process for identifying flare candidates. The black horizontal bar indicates the minimum flux required for each point in the flare candidate, and the red horizontal bar indicates the minimum flux required when including the contribution of the photometric error of each data point (shown as black error bars). The points in red were selected as the flare candidate following this procedure. The peak of the flare (vertical line) is taken to be the data point with the maximum flux. \label{fig:flarefinding}}
\end{figure}

Using the detrended light curve, flares can be identified as tranisent increases in flux, as shown in Figures~\ref{fig:detrending} and~\ref{fig:flarefinding}. We use  the flare finding algorithm by \citet{10.1088/0004-637x/814/1/35}, requiring that flares consist of at least six consecutive detrended flux values which are all at least 1.5 standard deviations above the mean (and 1.6 standard deviations when accounting for the photometric error of each data point). 

Detected flares may not necessarily have occurred on the target star -- flux contamination from a flare on an unresolved neighbouring star might also lead to flare signatures in the lightcurve. By measuring the flux-weighted average position (the `photometric centroid'; \citet{10.1086/671767}) over time, we can compare the measured centroid during a flare with the centroid when the star is not active. Large shifts in the centroid correlated in time with flares would provide evidence that a flare did not originate on the target star.

For each continuous observation period, we calculated the median centroid row and column of the star, and then subtracted this from the centroid time series. The standard deviation in the row and column centroid offsets were calculated, excluding datapoints marked as flares. Any flare candidates which exhibited an offset larger than 2 standard deviations in either their row or column centroids were marked for inspection.

For each flare candidate, we calculate the energy and estimate the full-width at half-maximum (FWHM) duration.  Energy calculations are done following the steps of \citet{10.3847/1538-3881/ab5d3a}, using temperature and radius estimates and uncertainties from the Tess Input Catalog (\citet{TIC}, TIC v8) if available, and GAIA-DR2 \citep{Gaia_Mission, Gaia_DR2} measurements otherwise. To estimate the FWHM, we fit the classical flare profile developed by \citet{10.1088/0004-637x/797/2/122} to each flare, and then calculate the FWHM using the fitted model. In some cases this is not possible, in which case the FWHM is estimated by fitting a second-order polynomial to the rise and decay phases separately. 

\subsection{Sample Completeness}

\begin{figure}
\plotone{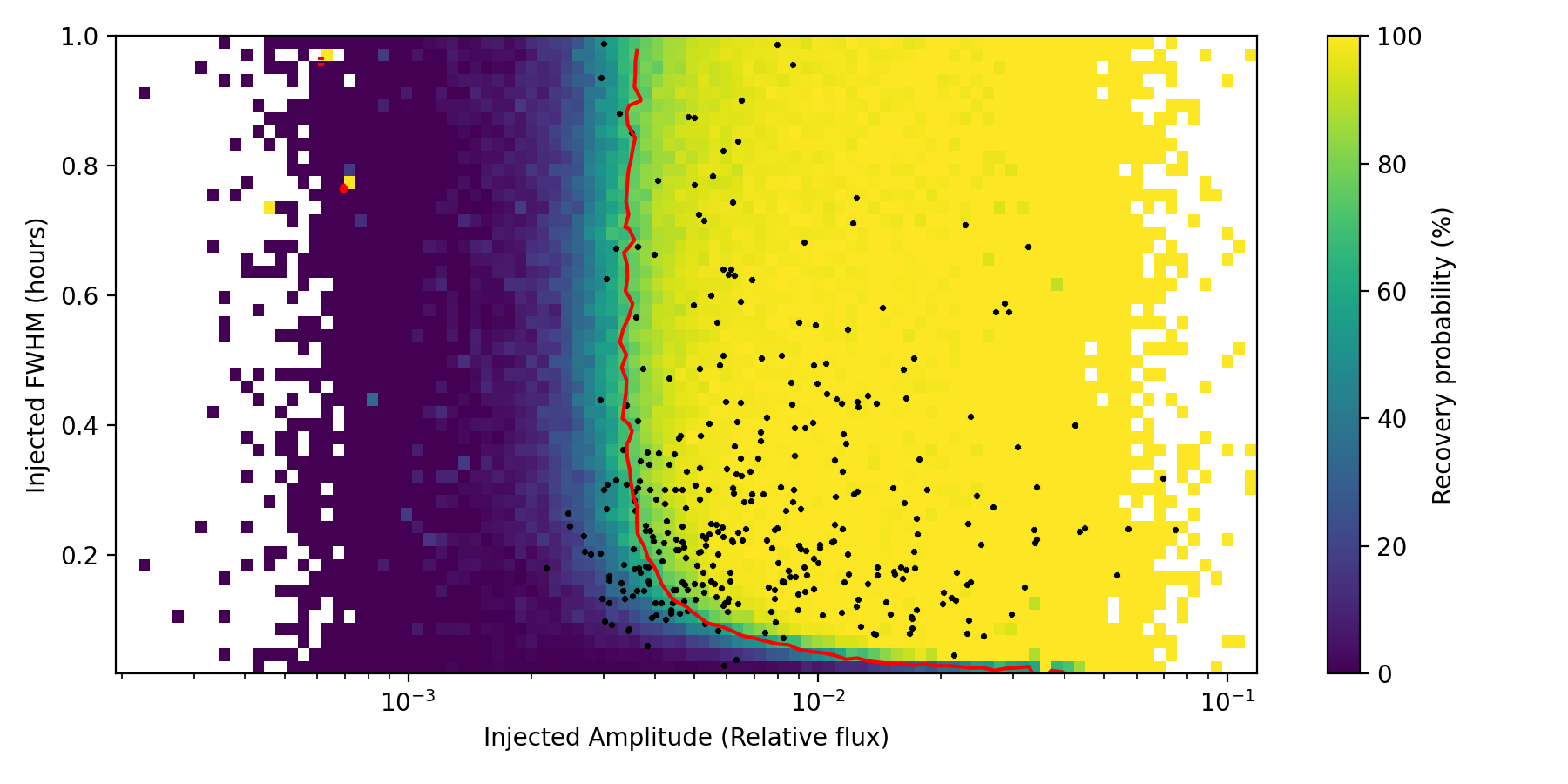}
\caption{
    The amplitude-FWHM-recovery rate surface of TIC 364588501 over all sectors, after 443,300 injected and recovered flares. The contour of 75\% recovery rate is shown in red. Observed flares are shown as black dots, however some observed flares had FWHMs above 1 hour. These flares are not shown.
    Sampling is incomplete towards the edges due to the use of a log-normal distribution of amplitudes.
    \label{fig:364588501_recov}}
\end{figure}

We test the robustness of our detrending and flare finding algorithms by injecting artificial flares into the light curve, and then recovering them. Using the classical flare profile from \citet{10.1088/0004-637x/797/2/122}, we generate flares with a specified amplitude and FWHM. The FWHM of each flare is drawn from a uniform distribution between 1 minute and 1 hour, and the amplitude of each flare is drawn from a log-normal distribution with the same mean and standard deviation as measured from the detected distribution of flares. 
We require the standard deviation of this log-normal distribution to be at least 0.3 relative flux units. For distributions with low variability (e.g.\ not many flares or flares with similar amplitudes) this ensures a more complete sampling of the amplitude-space. 
In practice, we find that a majority of observed flares on stars have estimated FWHMs within our sampled range, with the few flares with FWHMs longer than 1 hour having a large amplitude as well. 

For each continuous observation period, we inject 10 flares into the light curve and attempt to recover them using the detrending and flare-finding algorithms described in Sections~\ref{subsec:processing} and~\ref{subsec:flare-finding}. The generated flares are spaced with a buffer of 10 data points between any other flares, previously detected or previously injected, to avoid overlap. 
We considered an injected flare to be detected if the peak of any flaring candidate is within 10 data points of the time period where the injected flare was inserted.

After a sufficient number of injection and recovery cycles (at least 50,000 flares), we can estimate the recovery probability for flares based on their amplitude and FWHM. The recovery probability can be visualised in the form of a FWHM-amplitude-recovery probability surface, or as separate plots of the recovery probability as a function of FWHM or amplitude. Figure~\ref{fig:364588501_recov} shows an example of this recovery probability surface, and highlights the characteristic sharp decrease in the recovery probability at low amplitude and short FWHMs. 

For the individual recovery probability-amplitude and recovery probability-FWHM plots, we model the distributions using a logistic function. The choice of a logistic function follows from both the observed form of the distributions, and also from the expectation that at a sufficiently high amplitude or long FWHM, flares should always be recovered, and similarly at a sufficiently low amplitude or short FWHM the flare should be indistinguishable from noise. 

Using a fitted logistic function, we identify the amplitude at which the recovery probability is at least 75\%. All flares which are above this threshold are considered to be real flares and are marked for use in the analysis of the rate of flare occurrence. The FWHM of the flares is not factored into this filtering process -- we find that most flares which would be excluded from FWHM filtering are already removed due to the amplitude filtering.

\subsection{Rate Analysis} \label{sec:analysis}

Our statistical analysis focuses on identifying the rate of flaring for each star, and any variability in the rate. To achieve this we use the Bayesian Blocks algorithm \citep{10.1088/0004-637X/764/2/167}, which optimally determines a piecewise constant-rate model for a set of times of flare occurrence, consisting of a set of blocks (intervals with constant rate) and rates for each block. The two free parameter in the algorithm -- \texttt{ncp\_prior}, the prior for the number of change points, and \texttt{fp\_rate}, the percentage of detections that may be false positives -- are kept at their default values (\texttt{fp\_rate}$=0.05$, \texttt{ncp\_prior} calculated using Equation 21 from \citet{10.1088/0004-637X/764/2/167}). The flare data passed into the Bayesian Blocks data is treated as having no gaps in observation. This is achieved by calculating the length of each gap in observation, and shifting any time data after this gap by the length of the gap, essentially giving a measure of time in terms of the active observation time rather than in real time. 

To assign uncertainties to the Bayesian Blocks model we follow \citet{10.1088/0004-637X/764/2/167}, and perform a bootstrap analysis. We generate 1000 bootstrap samples of the flare data by sampling with replacement an equivalent number of flare event times. Then, the Bayesian Blocks algorithm is applied to each sample, and the rates and change points (the edges of each block) for each run are saved. Since both the number of blocks, their positions and their size can vary, we sample the rate of each run at 4096 equally spaced intervals in observation time. Then, we calculate the average (which we refer to as the `mean bootstrap rate') and the standard deviation (the `uncertainty') of the rates over the bootstrap runs at these intervals. 

The difference between the mean bootstrap rate and the optimal block representation gives an indication of how frequently that block is identified in the bootstrap samples, as well as information on when blocks are typically detected in the samples and their relative sizes. The uncertainty gives a measure of the spread in the rates -- how much the size of detected blocks varies between the bootstrap samples. 

It is of interest to investigate the waiting-time distributions (WTDs) for the flares. For each star, we generate the WTD for all flares, and also for flares in each of the identified blocks. In order to be consistent with the Bayesian Blocks algorithm, waiting times are calculated using the active observation time used by the flare data which the Bayesian Blocks algorithm is applied to, where gaps in observation are ignored. 
The WTD constructed in this way may differ from the actual WTD for flares from the star -- ignoring gaps in the data reduces longer waiting times which span the gaps, and can also introduce additional waiting times which are not present in the true distribution. For stars with a sufficiently high flaring rate (more than 2 flares per observation period), the constructed WTD should closely match the real WTD as the number of waiting times affected by ignoring gaps is only a small proportion of all the waiting times.

Following \citet{10.1023/A:1022430308641}, we use the Bayesian Blocks model to construct a model WTD for the flares, using equation~(\ref{eq:piecewise-poisson}). 

We also construct the flare frequency distribution (FFD) for each star. To investigate whether the FFD varies with flare rates for superflares, we also constructed the FFD for only flares in each of the blocks identified by the Bayesian Blocks algorithm. We fit a simple power-law model to each FFD using a Bayesian method with a uniform prior, which is equivalent to maximum likelihood estimation \citep{2010ApJ...710.1324W}. For some stars, there was a rollover at lower energies due to the incomplete recovery of lower energy flares, and for these stars we fit the simple power-law model above a minimum threshold energy, with the threshold being estimated by visual inspection of the distribution. 

We also examine the rotational phase distribution of flares. Previous analyses of the phase distributions focused on the overall phases of all flares on a star, however we investigate also whether there is a relationship between flare rate and phase dependence. Similarly to the FFD and WTD, we plot the phase distribution not only for the whole set of flares, but also for each block identified by the algorithm. We use the period from the Lomb-Scargle periodogram to fold the times for the star, with the position of 0 phase being arbitrarily chosen as the first data point in the time series. 

\section{Results} \label{sec:results}
\begin{deluxetable}{cccccccccc}
\tablenum{1}
\tablecaption{Parameters of sampled stars} \label{table:all_stars}
\tablehead{
\colhead{TESS ID \textsuperscript{a}} & \colhead{T$_{mag}$ \textsuperscript{a}} & \colhead{T$_{eff}$ \textsuperscript{a}} & \colhead{Mass \textsuperscript{a}} & \colhead{Radius \textsuperscript{a}} &\colhead{$\log g$ \textsuperscript{a}}& \colhead{Period}  & \colhead{Flares \textsuperscript{b}} & \colhead{$n_{sec}$ \textsuperscript{c}} & \colhead{$t_{obs}$ \textsuperscript{d}} \\
\colhead{} & \colhead{} & \colhead{(K)} & \colhead{(M$_\odot$)} & \colhead{(R$_\odot$)} &\colhead{} &\colhead{(days)} &\colhead{} &\colhead{} &\colhead{(days)}
}

\startdata
10202502 & 8.85  & 5351.0  & 0.93 & 0.88 & 4.52 & 3.54 & 2 & 2 & 44.94 \\
10357422 & 10.05 & 5215.0  & 0.89 & 1.07 & 4.33 & 1.35 & 3 & 3 & 70.65 \\
10747750 & 9.63  & 5310.0  & 0.91 & 0.82 & 4.57 & 5.12 & 5 & 5 & 108.44 \\
11642505 & 8.94  & 5382.0  & 0.93 & 0.84 & 4.56 & 6.15 & 2 & 1 & 24.78 \\
15862486 & 10.49 & 5807.63 & 1.04 & 1.00 & 4.46 & 5.39 & 1 & 1 & 23.16 \\
16146114 & 10.11 & 5324.73 & 0.92 & 0.80 & 4.59 & 5.71 & 1 & 1 & 24.04 \\
16331062 & 9.80  & 5606.0  & 0.99 & 1.05 & 4.39 & 2.26 & 2 & 3 & 46.93 \\
16493058 & 10.47 & 5605.04 & 0.99 & 0.68 & 4.77 & 1.40 & 1 & 1 & 18.19 \\
16878833 & 11.11 & 5256.01 & 0.90 & 0.72 & 4.68 & 1.86 & 7 & 2 & 31.33 \\
17201662 & 9.32  & 5601.91 & 0.99 & 0.91 & 4.51 & 3.93 & 1 & 1 & 24.43 \\
\enddata

\tablecomments{ 
\textsuperscript{a} Values from TIC v8 \citep{TIC}. 
\textsuperscript{b} Flare candidates above the 75\% recovery rate. 
\textsuperscript{c} Number of sectors observed by TESS. 
\textsuperscript{d} Total active observation length, excluding periods of observation less than 1 day long. \\
Excerpt of the full machine-readable table of the 270 stars analysed, showing the format of the data. 
}

\end{deluxetable}

\begin{deluxetable}{cccccccc}
\tablenum{2}
\tablecaption{Stellar parameters for stars exhibiting flaring rate variations.}\label{table:rv_stars}
\tablewidth{500pt}

\tablehead{
\colhead{TESS ID} & \colhead{T$_{eff}$ (K)} & \colhead{Mass (M$_\odot$)} & \colhead{Radius (R$_\odot$)} & \colhead{Period (days)} & \colhead{Flares} & \colhead{$n_{sec}$} & \colhead{Avg. Rate (days$^{-1}$)} 
}

\startdata
43472154    & $5316 \pm 122$    & $0.92 \pm 0.11$   & $0.90 \pm 0.05$       & $2.718\pm0.007$   & 50    & 2  & 1.01\\
85487971    & $5769 \pm 105$    & $1.03 \pm 0.13$   & $1.22 \pm 0.05$       & $2.703\pm0.008$   & 23    & 3  & 0.36\\
258771803   & $5701 \pm 127$    & $1.02 \pm 0.12$   & $1.07 \pm 0.05$       & $2.214\pm0.004$   & 26    & 18 & 0.076\\
284358867   & $5152 \pm 123$    & $0.87 \pm 0.11$   & $1.36 \pm 0.08$       & $3.303\pm0.012$   & 22    & 9  & 0.15\\
364588501   & $5605 \pm 135$    & $0.99 \pm 0.13$   & $1.22 \pm 0.06$       & $2.223\pm0.004$   & 280   & 26 & 0.46\\
382575967   & $5567 \pm 110$    & $0.98 \pm 0.13$   & $1.00 \pm 0.05$       & $2.195\pm0.004$   & 99    & 13 & 0.35\\
394030788   & $5231 \pm 120$    & $0.89 \pm 0.11$   & $0.90 \pm 0.05$       & $3.303\pm0.012$   & 282   & 12 & 0.97\\
\enddata

\end{deluxetable}

We analysed 274 G-type stars observed by \textit{TESS} for between a single sector (around 27 days) to 26 sectors (2 years; divided into 1 year periods of continuous observation). We excluded 4 stars which were not properly detrended -- these stars exhibited large periodic dips in their luminosity which were not properly modelled and led to erroneous flare detections. A total of 2703 superflares were detected on the remaining 270 stars, ranging in energy from $9.0 \times 10^{32}$ erg to $8.5 \times 10^{35}$ erg. A full machine-readable table of values is provided in Table \ref{table:all_stars}. We did not find any evidence of flux contamination via centroid measurements for any of our flares. This is likely due to the flare catalogues used having already removed stars which could be contaminated, e.g. binaries or unrelated bright neighbouring stars. 

We found that 16 stars in the sample had no detected flares and 72 stars had only 1 flare, despite our sample consisting of stars from catalogues of highly flaring stars. Manual examination of the light curves of these stars found no obvious flare-like structures, suggesting either any flares in the light curve were not being detected by our algorithm (e.g.\ they were very short-lived or small flares, or were the results of inaccurate de-trending), or potentially that there are erroneous flare detections in the catalogues used.

\begin{figure}
\plotone{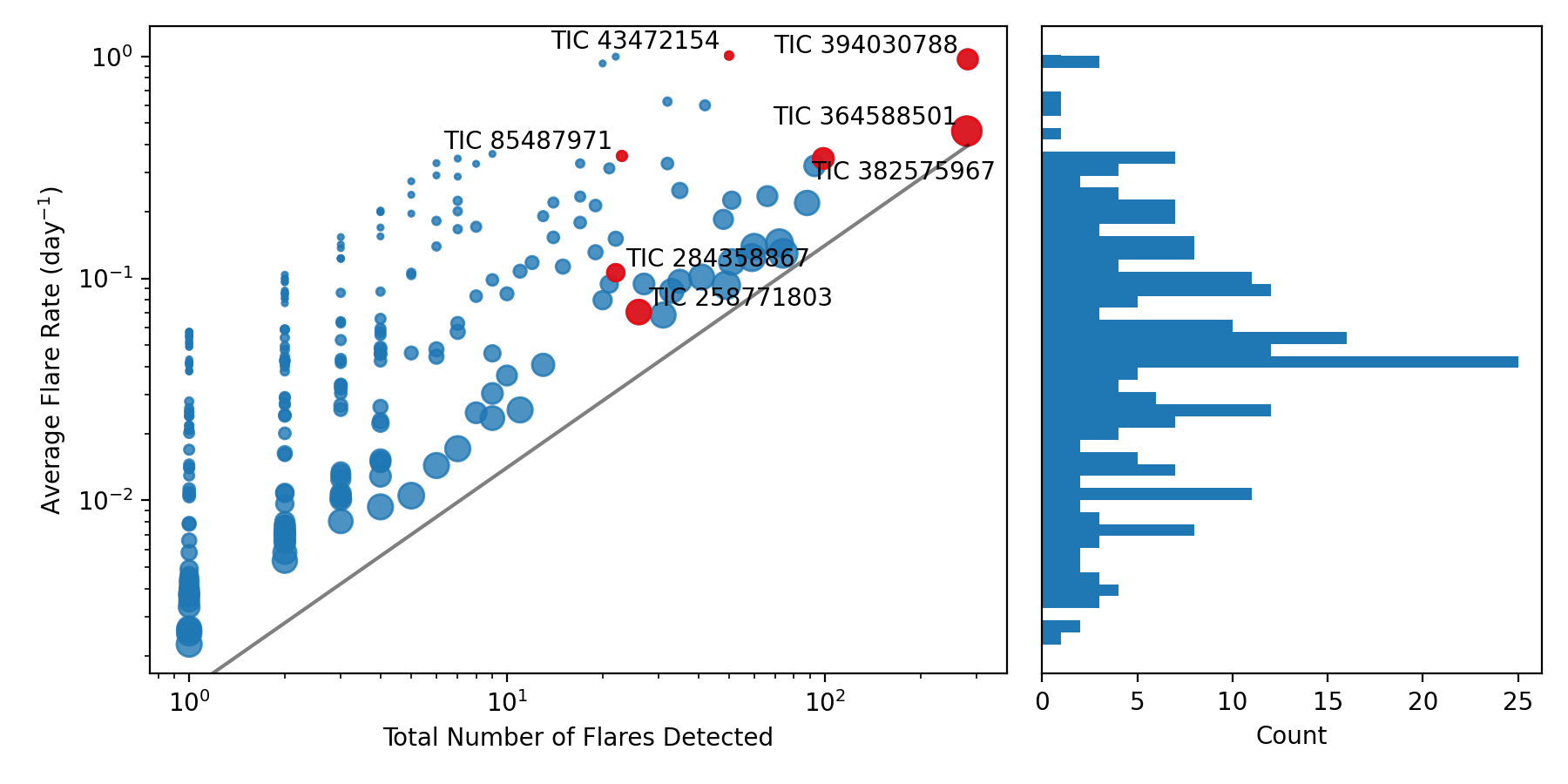}
\caption{
    Average flaring rate (total flares divided by observation time) of analysed stars versus the total number of flares detected. Stars with detected variations in their flaring rate are highlighted in red. The size of each marker is indicative of how many sectors the star was observed for. The lower limit (solid line) is the minimum possible flaring rate for a star, corresponding to observation for the maximum number of sectors possible (26). The histogram in the right panel shows the distribution of rates across the sample of stars.
    \label{fig:rate_vs_count}}
\end{figure}

We identified seven stars with variation in flaring rate, defined by having at least two different blocks identified by the Bayesian Blocks algorithm. These seven stars were confirmed to have real flaring events by manual inspection of their light curves. In Table \ref{table:rv_stars} we list the seven stars, together with their stellar parameters and the number of flares and sectors observed in each case. Rate variability was identified even in the cases of short observation times and low numbers of flares. We find no correlation between the rate variability of a star and the temperature or surface gravity ($\log g$). 

Figure \ref{fig:rate_vs_count} shows the average flaring rate for each of the 270 stars versus the number of flares detected. The seven stars with rate variation are shown in red, and the size of the circles indicates the number of sectors the star was observed for. The distribution of flaring rates is shown by the histogram at right. The figure shows that the three stars with the largest numbers of detected flares exhibit rate variation. 

\begin{figure}
\plotone{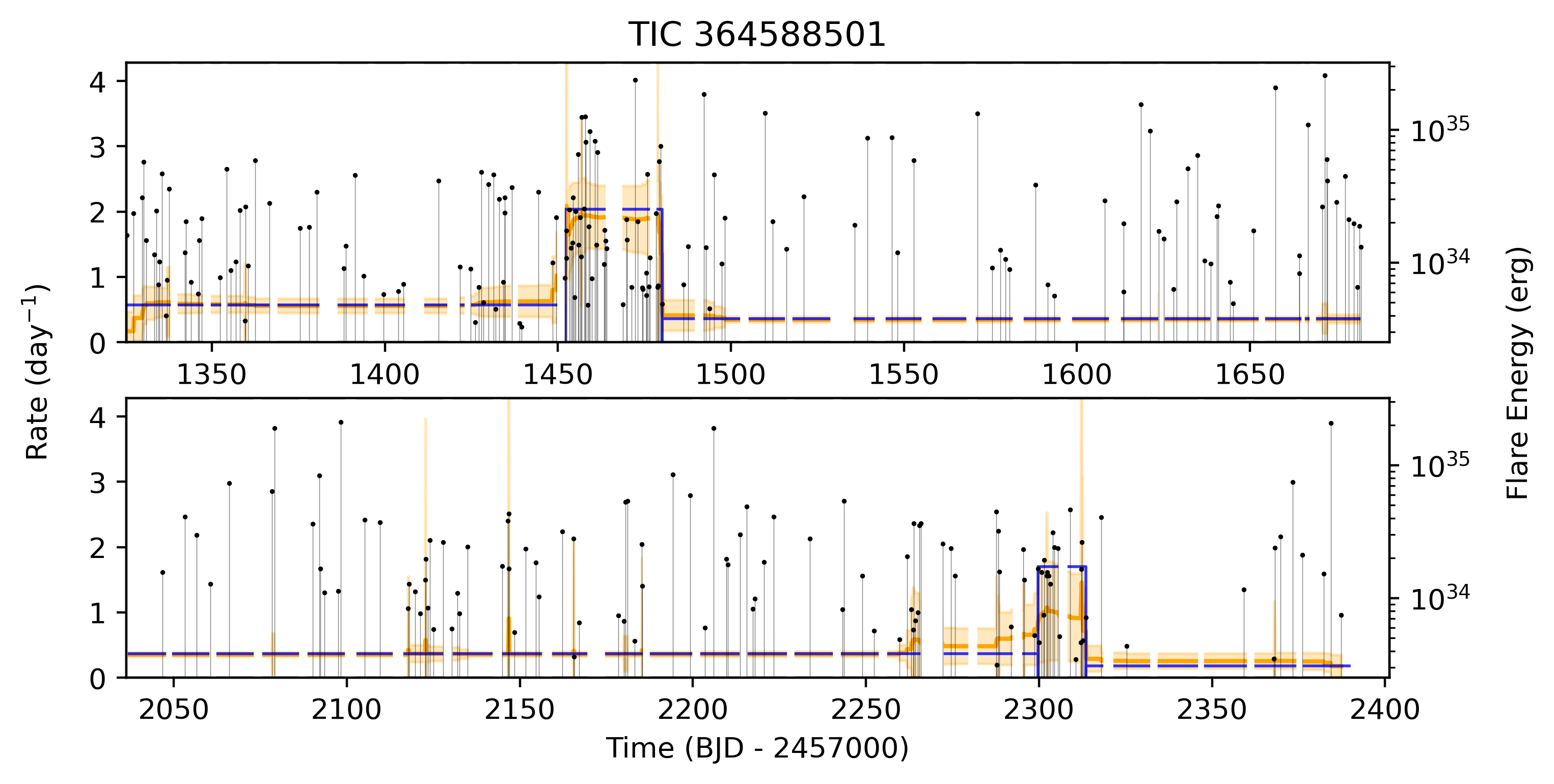}
\caption{
    Bayesian blocks determination of flaring rate (blue), bootstrap mean flaring rate and uncertainty (red and orange, respectively) and flares (black) for TIC 364588501 over years 1 (top) and 3 (bottom) of \textit{TESS} observations. The left-hand axis indicates the rates determined by the Bayesian Blocks and bootstrap methods, and the right-hand axis gives the flare energies.
    \label{fig:rateex1_1}}
\end{figure}

\begin{figure}
\plotone{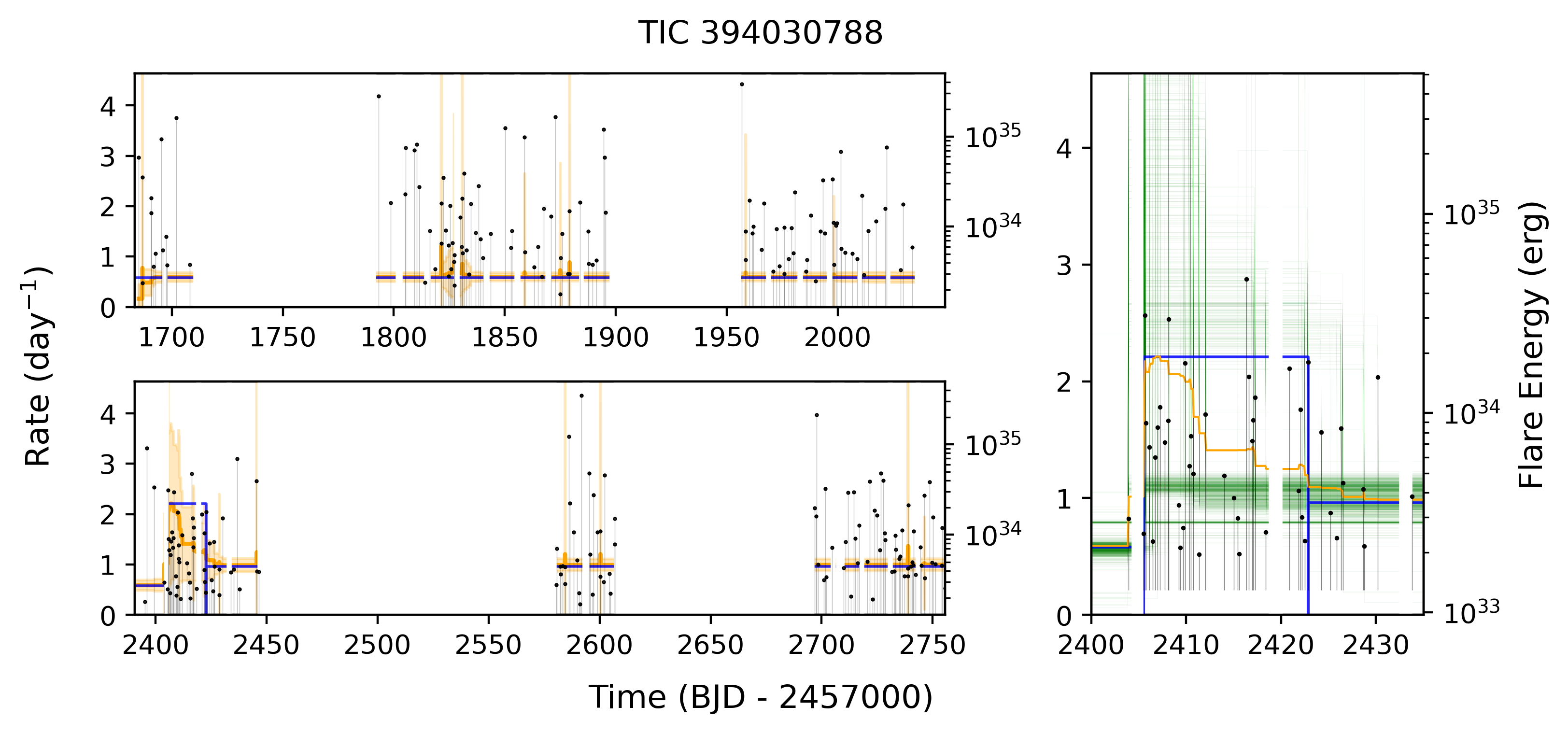}
\caption{
   \textbf{Left:} Bayesian blocks determination of flaring rate (blue), bootstrap mean flaring rate and uncertainty (orange and light orange, respectively) and flares (black) for TIC 394030788 over years 2 (top) and 4 (bottom) of \textit{TESS} observations.
   \textbf{Right:} The detected change in activity from the optimal block representation (blue), the bootstrap mean flaring rate (orange), flares (black) and the 1000 individual bootstrap samples (green).
    \label{fig:rateex1_2}}
\end{figure}

From the seven rate-variable stars, we found two distinct types of variability -- short term bursts of activity, and long-term changes in rate. The two stars with the highest number of detected superflares, TIC 364588501 and TIC 394030788, exhibited short bursts of high activity. 

Figure~\ref{fig:rateex1_1} shows the observed flares for TIC 364588501, together with the Bayesian Blocks and bootstrap analyses. Over its two years of observation with \textit{TESS}, two bursts of activity were identified by the algorithm. These bursts lasted for 28.0 and 25.7 days and increased the flaring rate by a factor of $3.6$ and $3.2$ respectively compared to the rate before each burst. 
Figure~\ref{fig:rateex1_2} shows the corresponding diagram for TIC 394030788. This star exhibited a similar burst of activity, with the rate increasing by a factor of $3.7$ for 17.3 days at around day 2400. 

From the shape and amplitude of the mean bootstrap rate, we can infer some general characteristics of the blocks identified in the bootstrap samples.
We found for both stars the mean bootstrap rate increased at the start of each identified block, indicative that the rate change was also identified to start around that time in the bootstrap samples. 
For TIC 364588501, the mean bootstrap rate also decreased sharply at the end of the identified blocks -- showing that most of the rate changes in the bootstrap samples in the block representation end around the same time as the end of the block in the optimal block representation. The first block of high activity on TIC 364588501 was also seen in both the mean rate and the optimal block representations, with the two rates nearly matching. 
However, for TIC 394030788 the decrease in rate at the end of the identified block occurred gradually, implying that the end time for the identified blocks varied between samples. In the right panel of Figure \ref{fig:rateex1_2}, we plot the block representation from each individual bootstrap sample in green. It can be seen that the duration and size of the blocks in each sample varies significantly, leading to the trailing off observed in the mean rate. 

\begin{figure}
\plotone{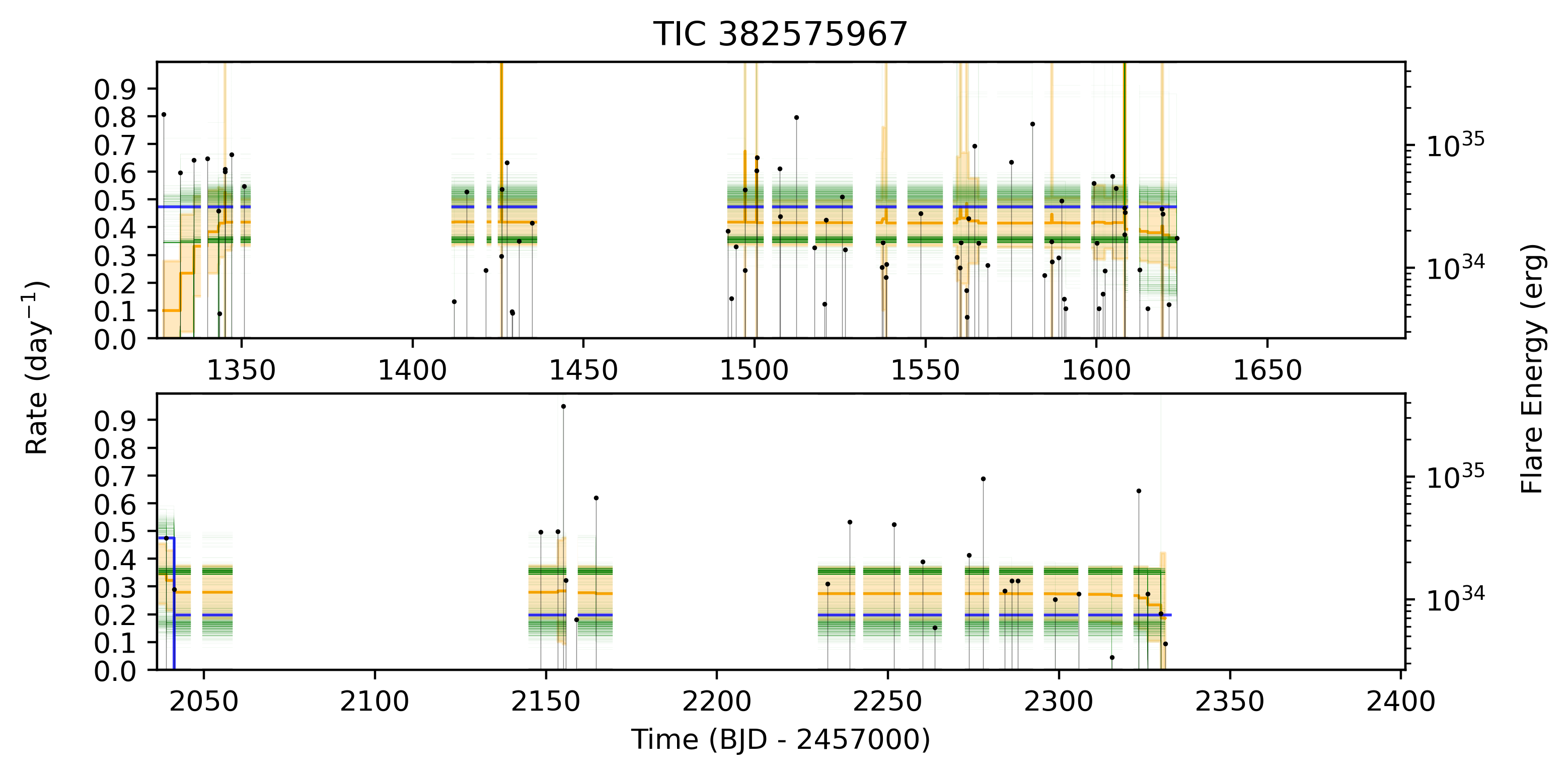}
\caption{
   Bayesian blocks determination of flaring rate (blue), as well as the individual bootstrap samples (green) for TIC 382575967 over years 1 (top) and 3 (bottom) of \textit{TESS} observations.
   \label{fig:rateex1_3}}
\end{figure}

For the other five rate-variable stars, only a single change in rate is identified by the algorithm. Visual inspection of the light curves generally confirmed the analysis: the higher activity interval of observation shows a notably higher number of flare detections compared to the lower activity interval. However, the mean bootstrap rate for these stars was nearly constant, only showing small increases at the locations of changes in rate identified by the optimal block representation. This implies that for a majority of the bootstrap samples, the optimal block fit was a single, constant rate.
In Figure \ref{fig:rateex1_3}, we show the optimal block representation, mean bootstrap rate and uncertainty, and the individual rates for each bootstrap sample for TIC 382575967. We observe only a small decrease in the mean bootstrap rate between the first and second identified blocks of activity, and can see that a large number of bootstrap samples imply a constant rate model.

\subsection{Flare Frequency Distributions}
We construct FFDs for each star using the whole set of flares, and then FFDs individually for each identified block from the Bayesian blocks algorithm. Uncertainties are calculated for both the energy estimates and for the counting statistics, the latter using Equations 11 and 14 from \citet{10.1086/164079} with a confidence limit of 1$\sigma$. 
We find that four of our rate variable stars have a rollover at lower energies, indicative of incomplete recovery of lower energy flares. This is expected due to the lower signal-to-noise ratio of these smaller flares due to their lower amplitudes and shorter durations. This is consistent with the recovery rates from the artificial flare injection and recovery procedure reducing with smaller amplitudes and durations. 
At high energies we also observe departures from a power-law. Similar breaks from the power-law distribution at high energies have been seen previously in G dwarfs \citep{10.3847/1538-4357/aafb76}.

\begin{figure}
\plotone{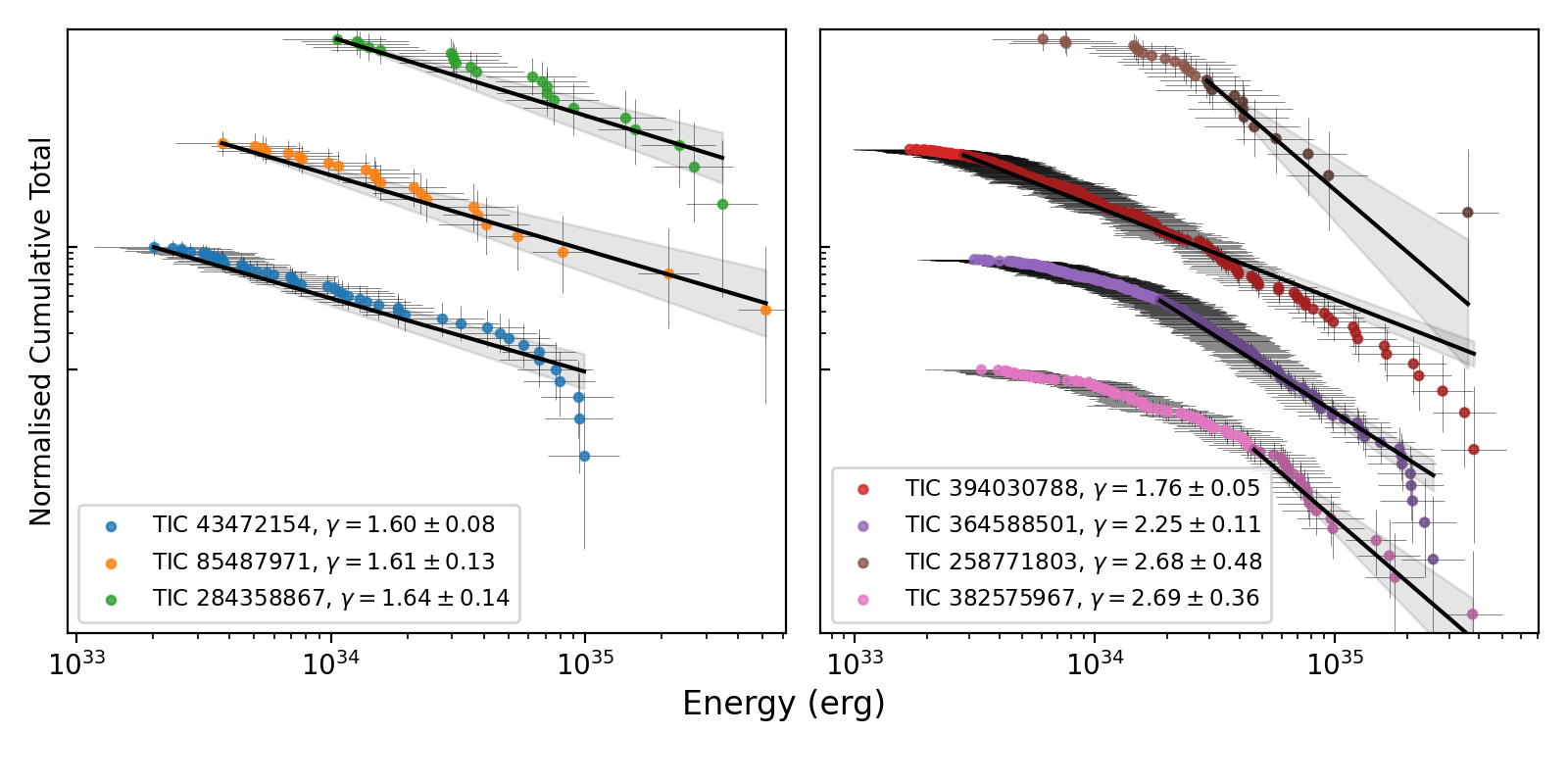}
\caption{
    \textbf{Left:} FFDs for the three rate-variable stars which were fitted using a simple power-law model, shown in log-log plots. Power-law fits are shown in black, with the uncertainty in the index shown in grey. Uncertainties in the bolometric energies of each flare are calculated following \citet{10.3847/1538-3881/ab5d3a}. Uncertainties in the count of each point are given by Poissonian counting errors following \citet{10.1086/164079}. The distributions for each star have been given arbitrary vertical offsets, to avoid overlapping.
    \textbf{Right:} Same, for the four rate-variable stars which were fitted using a simple power-law model with an energy threshold. Flares included in the fit are shown as darker circles. For each panel, the vertical axis represents the cumulative number of events with a larger energy, shown with a log scaling. 
    \label{fig:rvstars pl}}
\end{figure}

Figure~\ref{fig:rvstars pl} shows the FFDs for the seven stars which exhibited rate variation. A simple power-law model matched the distributions of three of the stars (TIC 43472154, TIC 85487971 and TIC 284358867), with the left panel showing the FFDs and power-law fits for these stars. The other four stars with rate variation (TIC 258771803, TIC 364588501, TIC 382575967 and TIC 394030788) exhibited significant departure from a simple power-law model, as shown in the right panel of Figure~\ref{fig:rvstars pl}. These stars were fit using the simple power-law model with a minimum energy threshold. 
The power-law indices for these stars are higher than those fit with a simple power-law distribution, and are high compared to indices for solar flare distributions (for total energy, $1.62\pm 0.12$, \citet{10.1007/s11214-014-0054-6}) as well as indices reported for superflares (e.g. $\gamma \approx 1.5$ \citep{10.1186/s40623-015-0217-z}, $\gamma = 1.76\pm 0.11$ \citep{10.3847/1538-4365/abda3c}). 
In the cases of TIC 258771803 and TIC 382575967, which have power-law indices $\gamma\approx 2.7$, the large indices may be due to the energy thresholds being too high and capturing the tail-end of the power-law distribution with a departure at high energies due to incomplete sampling from limited observation time. 

For the FFDs for each block, we attempted to fit the power-law models similarly to the FFDs for all flares. However, some of the distributions exhibit sharp departures from the power-law model. For the blocks we are able to fit successfully with power-laws, we find that the power-law indices for each block match the power-law index for the overall FFD. The fits for individual blocks have much larger uncertainties due to the lower number of flares.

\subsection{Waiting Time Distributions}
\begin{figure}
\plotone{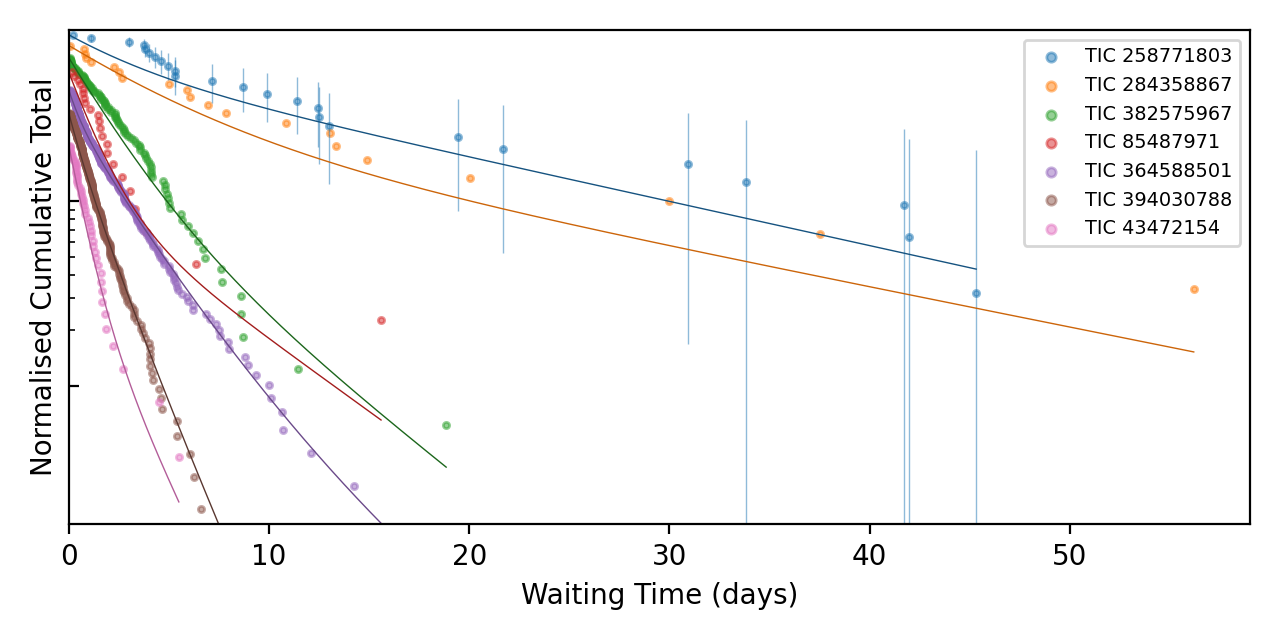}
\caption{
    The WTDs for all the rate-variable stars in our sample, shown in log-linear plots. The lines are the distributions corresponding to the Bayesian Blocks models. The distributions for each star have been given arbitrary vertical offsets to avoid overlapping.
    \label{fig:rvstars wtd}}
\end{figure}

Figure \ref{fig:rvstars wtd} shows the WTDs for the seven stars with rate variation, together with the piecewise-exponential model implied by the Bayesian Blocks results. There is a good qualitative agreement between the observed WTDs and the models, in particular for stars with more flares. 
 
\begin{figure}
\plotone{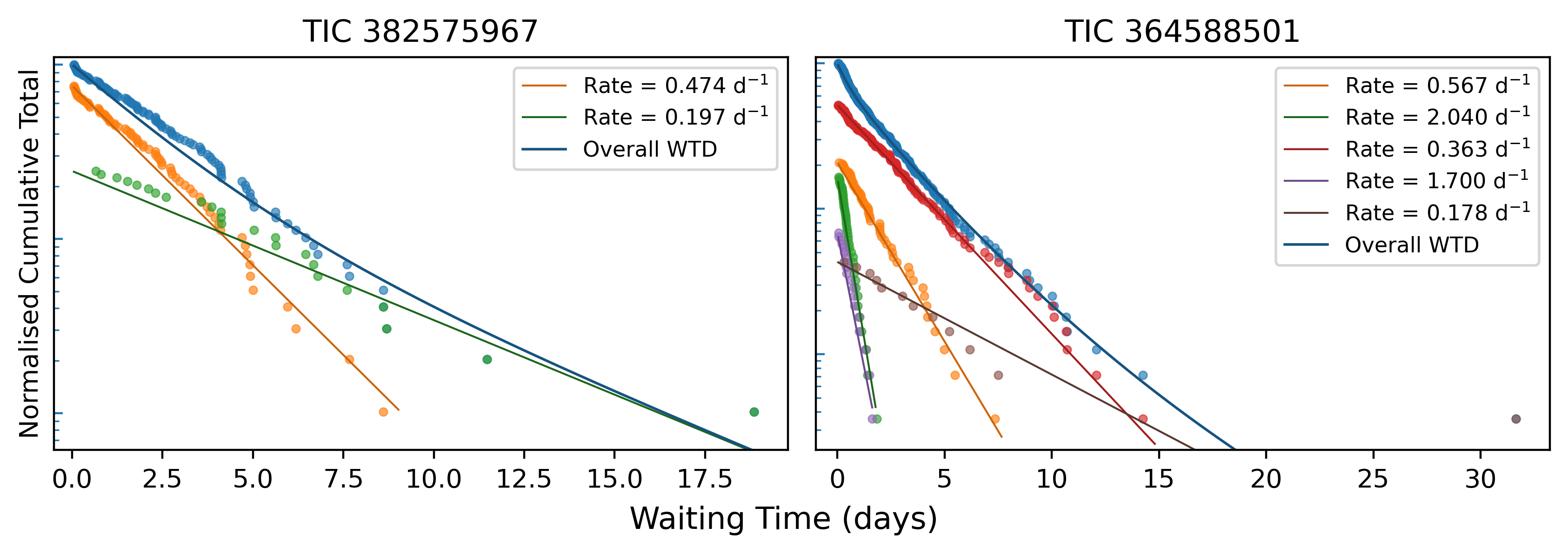}
\caption{
    Composite WTD log-linear plots, showing the WTD for the whole star (blue) and for each of the identified blocks (other colours), and the associated model for each distribution (coloured lines). For each panel, the vertical axis represents the cumulative number of events with a larger waiting time, shown with a log scaling. The distributions for the different stars have been given arbitrary vertical offsets, so that they do not overlap.
    \textbf{Left:} TIC 382575967, a star which exhibited only a single change in flaring rate.
    \textbf{Right:} TIC 364588501, a star which exhibited four rate changes. 
    \label{fig:wtd_composite}}
\end{figure}
 
The WTD for each of the identified blocks and the exponential (Poisson) models corresponding to that block also showed agreement for each of the stars. Figure~\ref{fig:wtd_composite} shows the WTD of all flares and the individual blocks for both TIC 382575967 and TIC 364588501. The exponential models for each of the identified blocks (shown as the non-blue distributions) are in good agreement with the distributions, particularly at shorter waiting times. On stars with smaller numbers of flares such as TIC 258771803 and TIC 284358867, deviations between the distribution and models can be attributed in part to the larger uncertainties associated with smaller numbers of flares. In Figure~\ref{fig:wtd_composite} we have indicated the uncertainties for TIC 258771803. These deviations from the piecewise-exponential model may also be attributed to the Bayesian blocks representation not capturing all rate variability in the data. This is particularly prevalent for stars with a smaller number of events, as the detection of rate changes is more difficult.

\subsection{Phase distributions}
\begin{figure}
\plotone{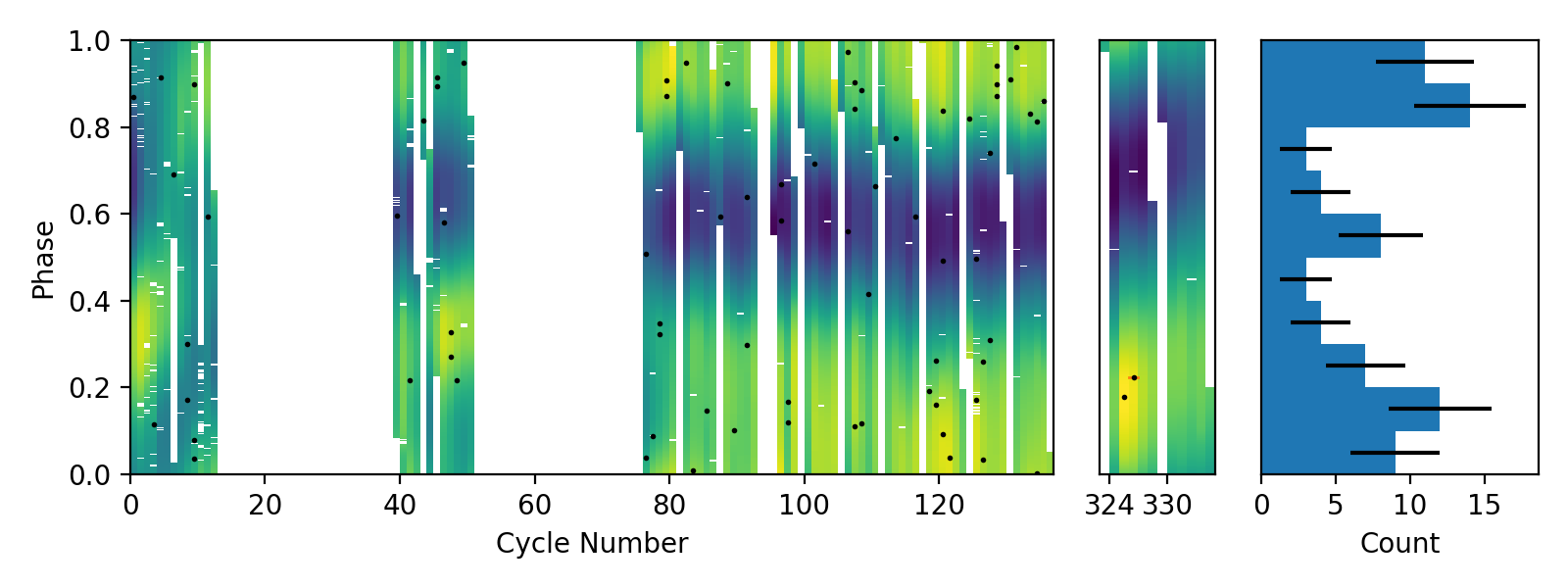}
\caption{
    The phase map of TIC 382575967 during one of its identified rate blocks. Colour indicates the amplitude of the quiescent luminosity model, and black dots indicate identified flares. The histogram on the right bins the flares according to their phase. Uncertainties in the histogram correspond to the square root of the counts.
    \label{fig:382575967_pb}}
\end{figure}

\begin{figure}
\plotone{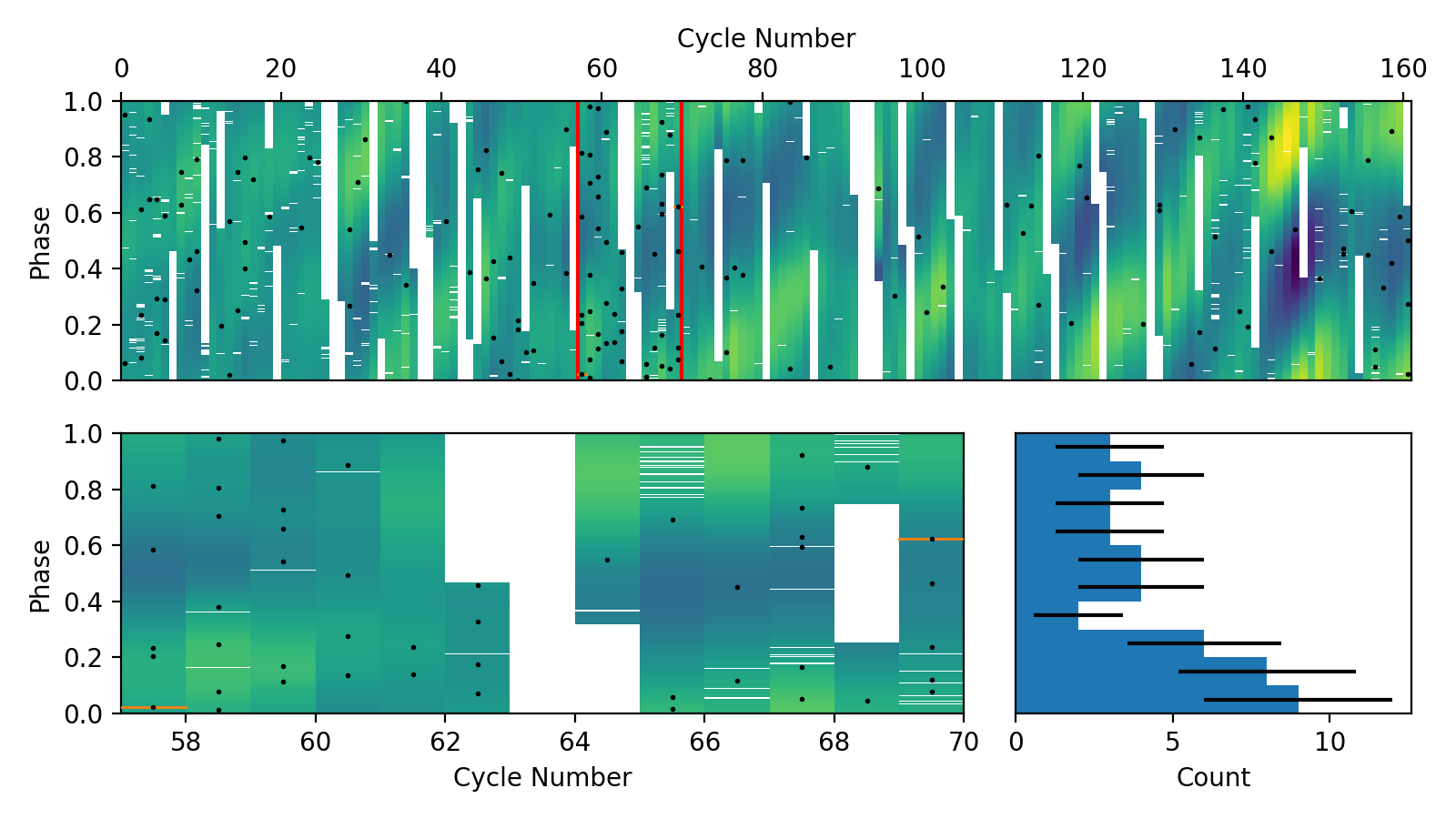}
\caption{
    \textbf{Top:} The phase map of the first year of \textit{TESS} observations of TIC 364588501. Flares are indicated as dots, and the colour indicates the amplitude of the quiescent luminosity model. 
    \textbf{Bottom:} Excerpt of cycles 57-69 (indicated in red on the above phase map), with the period of high flaring activity denoted between the two orange lines. The histogram on the side indicates the apparent phase dependence observed, using only the flares in the block. Uncertainties in the histogram correspond to the square root of the counts.
    \label{fig:364588501_pb1}}
\end{figure}

We calculate the phase distribution for the flares of each star (both the overall phase distribution and the distribution in each identified block). The phase is calculated by taking the modulus of the time using the period detected from the Lomb-Scargle periodogram. The flare phases are grouped into 10 bins and plotted as a histogram. For most stars, the distributions of flares with phase are roughly uniform, both overall and for each block. However, TIC 364588501 and TIC 382575967 show possible phase dependence, as seen in Figures~\ref{fig:382575967_pb}~and~\ref{fig:364588501_pb1}. 

Our choice of phase is chosen arbitrarily based on the first data point, however we can further investigate the phase dependence of flares and any evolution in the rotational modulation by plotting the background flux model folded over each rotational cycle using the period detected with the periodogram. We refer to these diagrams as phase maps. These phase maps are similar to the diagrams used by \citet{10.1088/0004-637X/806/2/212} and \citet{10.3847/1538-3881/ab9536}. On these phase maps, we also plot the positions of flares in terms of their phase and their rotational cycle.

Figure \ref{fig:382575967_pb} shows the phase map for TIC 382575967, and reveals a consistent darkening across many cycles at the same phase. This is indicative of a persistent structure on the stellar surface rotating with the same period as measured by the periodogram. The times of flares are indicated by the dark spots on the phase map, and the phase histogram for the flares is the right-hand panel. The histogram indicates some tendency for flares to occur at a phase corresponding with maximum brightness in the rotational modulation. 
Comparison of the phase distribution to a uniform distribution using a Kolmogorov-Smirnov (K-S) test gave a $p$-value of 0.076. This $p$-value is very low compared to other phase distributions, and provides some evidence that the phase distribution is not well fit with a uniform distribution.

For several stars (including TIC 364588501) the histogram plots did not match up with the expected phase distribution when checking the light curve directly. We observed for these stars that the rotational modulation occurred with a different periodicity than the dominant periodicity detected through the periodogram, and this was evident through the phase map. 
Structures on the surface which are not rotating with the same periodicity as the dominant period used for folding will appear to shift forward or backwards in phase in each consecutive cycle. This leads to diagonal streaks in the phase map. For particularly stable structures, it is possible to fit a linear model to calculate the rotational period \citep{10.1088/0004-637X/806/2/212, 10.3847/1538-3881/ab9536}.

\begin{figure}
\plotone{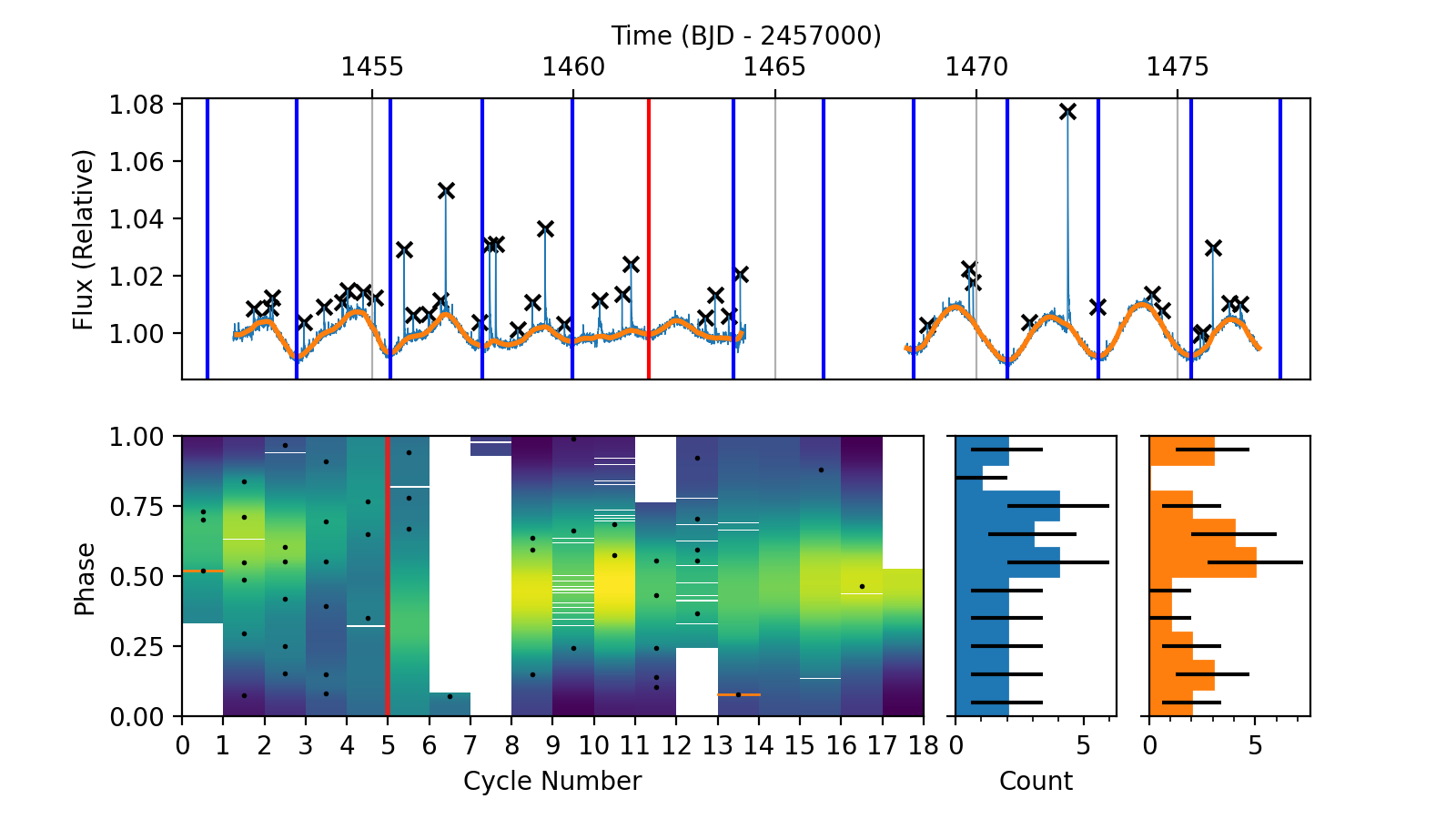}
\caption{
    \textbf{Top:} Excerpt of light curve of TIC 364588501 of the first identified block of high flaring rate on the star, showing the relative flux (blue), quiescent luminosity model (orange), flares (crosses) and identified periods using minima from the model (blue lines). The star initially has complicated oscillations, however starting with the fifth manually identified cycle (red; cycle 62 overall) the rotational modulation shows a more regular pattern.
    \textbf{Bottom:} Adjusted phase map and phase distribution for the whole first block of high flaring rate, using the identified periods as above. The phase distribution is split between flares before cycle 5 (blue) and after cycle 5 (orange). 
    \label{fig:364588501_pb2}}
\end{figure}

Figure \ref{fig:364588501_pb1} shows the phase map for TIC 364588501. During the identified block between cycles 57 and 69, the flare phase distribution histogram suggests a possible dependence on phase. 
However, in the top panel for Figure \ref{fig:364588501_pb2}, we show an excerpt of the light curve during this block. It can be seen that unlike TIC 382575967, the oscillations in the light curve do not match the folding period, resulting in diagonal streaks. 
Visual analysis of the light curve shows a complicated oscillation pattern up to cycle 62 which appears to be two out of phase sinusoids with different amplitudes. By the end of cycle 62, one of these sinusoidal oscillations has decayed, and starting with cycle 64 we see a clear single sinusoidal oscillation. We interpreted this as two groups of starspot structures, with one decaying by cycle 62.

In order to measure the phase for each flare with respect to the observed oscillations, we first identify the start and end of each cycle. In Figure~\ref{fig:364588501_pb2}, we show an excerpt of the light curve. Each cycle is identified by locating a minimum in the quiescent luminosity model, and then stepping forward by the measured period and identifying the minimum in a one day window around this time. For cycles at the start and end of a period of continuous observation, we assume their length is the measured period. The length of each cycle identified in this way varies, with the cycles having an average duration of $2.38 \pm 0.55$ days. Calculating the phase of each data point and flare using these identified cycles gives the phase map shown in Figure~\ref{fig:364588501_pb2}. We calculate the phase distribution separately for the flares before and after the fifth manually identified cycle, corresponding to the aforementioned change in the oscillation pattern. Individually, each phase distribution appears to be roughly uniform, with the phase distribution of the starspot structure after the fifth cycle (shown in orange) showing a possible preference to flaring towards the peak of the oscillation. 
We individually compared each of the distributions to a uniform distribution using a K-S test. For both distributions, the $p$-values from the K-S test were high ($p$-values of 0.93, 0.53 for before and after the fifth cycle, respectively), suggesting that both of these phase distributions can be fit well with a uniform distribution.

\section{Discussion} \label{sec:discussion}
We were able to identify seven stars out of our sample of 274 G-type stars which exhibited a change in their rate of superflare occurrence using the Bayesian Blocks algorithm. The number of flares detected and average flaring rate varied significantly between these rate-variable stars, with rate variability being identified with as few as 21 flares on TIC 284358867. We were also able to qualitatively evaluate the uncertainty in the piecewise constant rate model and, by extension, the presence of rate variability in these stars through a bootstrap analysis.

We saw broad agreement between the FFD and WTD and their respective models, highlighting the similarities in the statistics of superflares and solar flares. However, our measured power-law indices for the FFDs were notably higher than similar measurements for solar flares ($1.62\pm 0.12$, \citet{10.1007/s11214-014-0054-6}) and other superflare papers. For two of our rate variable stars which had high power-law indices around $\gamma \approx 2.7$, we attributed this to improper detection of the position of the rollover seen at lower energies due to incomplete recovery of flares. However, two other of our rate-variable stars with rollovers and many non-rate variable stars had power-law indices around $\gamma \approx 2.0$, which is still appreciably higher than solar flare and other superflare measurements.

Given the relationship observed between solar flares and sunspots, it is unusual that rate variability was identified on only a small number of stars in our sample. Most of the stars analysed in our sample exhibited rotational modulation, which is typically attributed to the presence of starspots \citep{10.1088/0004-637X/771/2/127} (and has been confirmed through spectroscopic measurements \citep{10.3847/1538-4357/ab14e6}). Despite this, many stars exhibited no rate variability, and even among the few stars which were identified as rate variable we only found one convincing case where flares appeared to exhibit some phase preference on TIC 382575967.
Uniformity in the phase distribution of stellar flares has previously been seen in other studies \citep{10.1088/0004-637x/797/2/121, 10.1093/mnras/sty1963, 10.1093/mnras/stz2205, 10.1093/mnras/staa923}, and was attributed to either flaring across multiple active regions or in a polar starspot/starspot group which is always in view. The inability to spatially resolve starspots makes discerning between these two scenarios difficult. If a star is less spotty, has fewer active regions or flaring is predominantly caused by the dominant starspot group, this could lead to the phase preference seen on TIC 382575967.

The presence of multiple flaring active regions on the stellar disk of each star should not affect the WTD if flaring occurs in each active region as a Poisson process. 
On the Sun, flaring in active regions have been noted to be approximately a Poisson process; exhibiting a constant flaring rate (at least for short periods of observation). The superposition of multiple Poisson processes gives another Poisson process with a rate given by the sum of all the comprising processes' rates. This is seen on the Sun, where the whole-Sun WTD is approximately exponential over short timescales (as in Equation~\ref{eq:piecewise-poisson}), and for individual active regions where the WTD is exponential. 
We observe on superflaring stars that the overall WTD is well fit by Equation~\ref{eq:piecewise-poisson}, and the WTD of each identified constant-rate blocks by the Bayesian Blocks algorithm is also exponential. This suggests that flaring in each of these active regions is a Poisson process, and that we cannot distinguish between individual active regions through distributions such as the WTD -- a single active region and multiple active regions are statistically identical.

The low number of stars with rate variability may also be attributed to systematic issues on some stars, such as a low number of detected flares or limited observation time. 184 stars in our sample had an average flaring rate of less than 1 flare per 14 days (the average length of an observation period), and 226 stars had less than 10 flares after filtering out lower energy flares. For stars with low flaring rates, the limited observation time of \textit{TESS} is the primary limit, with some stars only having 10-20 detected flares even after two years of continuous observation. While we were able to detect rate variability with similar flare counts (TIC 284358867 with 21 flares), the detected rate changes are more uncertain based on the bootstrap analysis -- often we find for stars with lower flare counts many bootstrap samples are still best fit with a constant model. In one case, the star TIC 236757919 was detected as rate variable, but detection of additional flares from a new sector of data resulted in the optimal block representation changing to a constant-rate model. 

Measuring flaring rates directly is difficult, as every flare contributes equally to distributions such as the WTD and to rate calculations, independent of flare size. Due to the incomplete recovery of lower energy flares, we are forced to choose between introducing biases by including lower energy flares with lower recovery rates (potentially missing some data), or limiting our dataset further. Other methods of quantifying flaring activity, such as the fractional flux emitted by flares \citep{10.3847/2515-5172/ab45a0,10.3847/1538-3881/ab9536} or the fraction of time spent flaring \citep{10.3847/1538-4365/aab779} avoid this issue by implicitly limiting the contribution of these harder-to-detect flares, and have seen some success in identifying longer time-scale variability. However on shorter time-scales of days or weeks, these measurements are noisy due to the wide range of energies of flares. 

With the \textit{TESS} mission ongoing, additional observations from the next few years will help improve our analyses of these stars. 
Further analysis of short-term rate variability could be performed with M dwarf stars, which are usually more active than G dwarfs and are easier to detect flares on due to their lower luminosity. In future studies we will also use \textit{TESS} observations of stars with known activity cycles to attempt to detect rate variability over longer time scales.

\begin{acknowledgments}
This research is supported by an Australian Government Research Training Program (RTP) Scholarship.
\end{acknowledgments}

\bibliography{paper}{}
\bibliographystyle{aasjournal}

\end{document}